\documentclass[12pt, a4paper]{article}
\usepackage[utf8]{inputenc} 
\usepackage{chet}
\usepackage{authblk}
\usepackage{slashed}
\usepackage{indentfirst}
\usepackage{amsmath}
\usepackage{amsfonts}
\usepackage{amssymb}
\usepackage{courier}

\title{Constructing massive superstring vertex operators from massless vertex operators using the pure spinor formalism}
\author{Bruno Rodrigues Soares\footnote[1]{\texttt{br.soares@unesp.br}}}
\affil{\textit{ICTP South American Institute for
Fundamental Research}\\ \textit{Instituto de Física Teórica, UNESP - Univ. Estadual Paulista}\\
\textit{Rua Dr. Bento Teobaldo Ferraz 271, 01140-070, São Paulo, SP, Brazil}}
\abstract{The vertex operator for the first
massive states of the open superstring is constructed in terms of d=10 super Yang-Mills superfields using the OPE's of massless vertex operators in the pure spinor formalism.}

\begin{document}
\maketitle
\section{Introduction}
\label{Introduction}

The pure spinor formalism for the superstring has all spacetime symmetries manifest \cite{Berkovits:2000fe}. This feature allows the construction of super-Poincaré covariant expressions for vertex operators through its quantization \cite{Berkovits:2002qx, Berkovits:2000yr}. These operators correspond to physical states in the cohomology \cite{Berkovits:2000nn} of the BRST charge $Q = \oint dz \lambda^\alpha d_\alpha$, which is expressed in terms of a ten-dimensional worldsheet spinor $\lambda^\alpha$ satisfying the pure spinor condition and the worldsheet variable $d_\alpha$ for the space-time supersymmetric derivative. The knowledge of vertex operators makes it possible to establish the equivalence of superstring amplitudes in the pure spinor and RNS formalisms \cite{Berkovits:2005ng}. Nevertheless, a superfield description of superstring massive vertex operators remains an open problem.

In order to construct open superstring unintegrated vertex operators of mass $m^2 = 2n$, one can write every possible combination of worldsheet fields with ghost number 1 and conformal weight $n$, and contract them with d=10 superfields. The onshell condition provides relations between these d=10 superfields \cite{Berkovits:2002qx}. 
In the case of integrated operators, one needs to use the descent relation to constrain the d=10 superfields \cite{Chakrabarti:2018mqd}. Although straightforward, this method becomes quite involved at higher mass levels and it is convenient to resort to other ways of building the corresponding vertex operators.


In this paper, the open string unintegrated vertex operator at the first massive level will be constructed from the operator product expansion between a massless integrated and a massless unintegrated vertex operator using pure spinor formalism CFT. This massive vertex will be BRST invariant by construction and expressed in terms of super Yang-Mills d=10 superfields which have well-known theta expansion \cite{1986CMaPh.106..183H}. This result can be generalized for any higher mass level and used to compute scattering amplitudes with massive vertex using all the machinery known for massless scattering amplitude computations \cite{Kashyap:2023cdi}.

In section \ref{sec2}, after a brief review of pure spinor formalism, the unintegrated vertex operator at the first mass level will be computed, and its BRST invariance will be verified. In section \ref{sec3} the gauge symmetries will be used to find a gauge where the vertex operator superfields are related with the usual supergravity superfields \cite{Berkovits:2002qx}.

Note: While this work was being completed, the paper \cite{Kashyap:2023cdi} appeared which contains the main results discussed here as well as other results on massive amplitudes. However, the work here presents computations which were performed independently and were not included in \cite{Kashyap:2023cdi}. After completing this work, the authors of \cite{Kashyap:2023cdi} have informed me that they have also performed similar computations which will soon be posted on the arXiv together with further results on massive amplitudes.

\section{Massive Vertex Operator} \label{sec2}

The pure spinor formalism for the open string has the following action
\begin{equation}
    S_{PS} = \frac{1}{\pi} \int d^2z \big( \frac{1}{2} \partial x^m \bar{\partial} x_m + p_\alpha \bar{\partial} \theta^\alpha - w_\alpha \bar{\partial} \lambda^\alpha  \big),
\end{equation}
where $m = 0,...,9$, and $\alpha = 1,...,16$ are the vector and spinorial indices of $SO(10)$, together with a nilpotent BRST operator 
\begin{equation}
    Q = \oint dz \lambda^\alpha d_\alpha, \label{brstcharge}
\end{equation}
with the GS constraint defined as $$d_\alpha = p_\alpha - \frac{1}{2} \partial x^m (\gamma_m \theta)_\alpha - \frac{1}{8} (\theta \gamma^m \partial \theta) (\gamma_m \theta)_\alpha, $$ and the field $\lambda^\alpha$ satisfying the pure spinor property $\lambda^\alpha \gamma^{m}_{\alpha \beta} \lambda^\beta = 0$. The worldsheet variables $\theta^\alpha$, $\lambda^\alpha$ have conformal weight $h=0$ and their conjugate pairs $p_\alpha$, $w_\alpha$ have conformal weight $h=1$. There is a ghost current $ J= w_\alpha \lambda^\alpha $ that can be used to define the ghost number of pure spinor operators. 

The integrated and unintegrated vertex operators are \cite{Berkovits:2000fe}
\begin{align}
    \label{integrated_massless}
    U(z) &= \  :\Pi^m A_m : + :\partial \theta^\alpha A_\alpha: + :d_\alpha W^\alpha: + :\frac{1}{2} N^{mn} F_{mn}:,  \\
    V(z) &= \lambda^\alpha A_\alpha ,\label{unintegrated_massless}
\end{align} with supersymmetric momentum $\Pi^m = \partial x^m + \frac{1}{2} (\theta \gamma^m \partial \theta)$, the Lorentz current $N^{mn} = \frac{1}{2} w \gamma^{mn} \lambda $ and superfields $[A_m$,$A_\alpha$,$W^\alpha$,$F_{mn}]$ built out of $A_{\alpha}$,
\begin{align}
    W^\alpha &= \frac{1}{10} (\gamma^m)^{\alpha \beta} \big( D_\beta A_m - \partial_m A_\beta \big) \label{sym_eom:a}\\
    A_m &= \frac{1}{8} \gamma_m^{\alpha \beta} D_\alpha A_\beta \label{sym_eom:b}\\
    F_{mn} &= \frac{1}{8} (\gamma_{mn})_{\ \beta}^{\alpha} D_\alpha W^\beta,  \label{sym_eom:c}
\end{align} and their super Yang-Mills equations implies the onshell condition $ Q \cdot V = 0$ and the descent relation $Q \cdot U = \partial V$. The normal ordering $: \cdot :$ prescription is defined as \cite{DiFrancesco:639405}
\begin{equation}
    :A(z) B(w): \ \equiv \oint \frac{dz}{z-w} A(z) B(w). 
\end{equation}

The relevant OPE's for subsequent computations are
\begin{align}
    x^{m}(z, \bar{z}) x^{n}(w, \bar{w}) &\sim -\delta^{m n} \ln |z-w|^{2}, &d_{\alpha}(z) \theta^{\beta}(w) &\sim \frac{\delta_{\alpha}^{\beta}}{z-w},\nonumber \\
    d_{\alpha}(z) d_{\beta}(w) &\sim -\frac{\gamma_{\alpha \beta}^{m} \Pi_{m}(w)}{z-w}, \ &\Pi^{m}(z) \Pi^{n}(w) &\sim -\frac{\delta^{m n}}{(z-w)^{2}},  \nonumber \\
     d_{\alpha}(z) \Pi^{m}(w) &\sim \frac{\left(\gamma^{m} \partial \theta(w)\right)_{\alpha}}{z-w}, \ &\Pi^m(z) V(w) &\sim - \frac{\partial^m V(w)}{z-w} , \nonumber \\
     N^{m n}(z) \lambda^{\alpha}(w) &\sim  \frac{1}{2} \frac{\left(\gamma^{m n}\right)^{\alpha}_{\ \beta} \lambda^{\beta}(w)}{z-w},  &d_\alpha(z) V(w) &\sim D_\alpha V, \label{OPES}
\end{align}
where $V(w) = V(\theta) e^{ik\cdot x}$ is a superfield, $D_\alpha = \frac{\partial}{\partial \theta^\alpha} + \frac{1}{2} (\gamma^m \theta) \partial_m$ is the supersymmetric derivative, and $\partial_m \equiv \frac{\partial}{\partial x^m}$.

The operator algebra of string theory primary fields can be used to recover all theory higher mass resonances \cite{Friedan:1985ge}. Since unintegrated vertex operators of mass $m^2 = 2n$ should be constructed from combinations of $[\Pi^m$, $d_\alpha$, $\theta^\alpha$, $N^{mn}$, $J$, $\lambda^\alpha]$ with ghost number $1$ and conformal weight $n$, one can define the unintegrated vertex operator corresponding to the first massive state as
\begin{align}
    V^{(12)}_{m^2 = 2} \equiv  \oint d z_1 \ U^{(1)}(z_1)V^{(2)}(z_2), \label{claim} \\ (k_1 + k_2)^2 = -2, \label{v12momenta}
\end{align}
where $U^{(1)}$ and $V^{(2)}$ are integrated and unintegrated massless vertex operators, respectively. The onshell and descent relations of $V^{(2)}$ and $U^{(1)}$ implies $Q\cdot V^{(12)}_{m^{2} = 2} = 0$. 

To write \ref{claim} in terms of super Yang-Mills superfields, first consider the OPE between the first term of \ref{integrated_massless} with \ref{unintegrated_massless},
\begin{align}
    :\Pi^m A^1_m (z_1): :\lambda^\alpha A^2_\alpha(z_2): 
    = &+\frac{1}{z_{12}}  :\Pi^m A^1_m \lambda^\alpha A^2_\alpha(z_2) : -\frac{1}{z_{12}} :  \partial A^1_m \lambda^\alpha \partial^m A^2_\alpha (z_2):. \label{first_term_answer}
\end{align}
Using the equation $\partial K = \partial \theta^\alpha D_\alpha K + \Pi^m (ik_m) K$ on \ref{first_term_answer}, one has
\begin{align}
     \oint_{z_2} d z_1 :\Pi^m A^1_m (z_1): \lambda^\alpha A^2_\alpha (z_2) &= \ :\Pi^m \lambda^\alpha  A^1_m A^2_\alpha: \nonumber \\ &\ - :\partial \theta^\beta \lambda^\alpha D_\beta A^1_m  \partial^m A^2_\alpha:  \nonumber \\ &-  :\Pi^m \lambda^\alpha (ik^1_m) A^1_n  \partial^n A^2_\alpha: \ .
\end{align}
Considering the other terms of \ref{integrated_massless}, one obtains
\begin{align}
     \oint_{z_2} d z_1 :\partial \theta^\beta A^1_\beta (z_1): \lambda^\alpha A^2_\alpha (z_2) &= \ :\partial \theta^\beta \lambda^\alpha A^1_\beta A^2_\alpha:,
\end{align} 
\begin{align}
     \oint_{z_2} d z_1 :d_\beta W^{1 \beta} (z_1): \lambda^\alpha A^2_\alpha(z_2) &= : d_\beta  \lambda^\alpha W^{1 \beta} A^2_\alpha : \nonumber \\ &- : \partial \theta^\beta \lambda^\alpha D_\beta W^{1 \xi} D_\xi A^{2}_\alpha  : \nonumber \\ &- : \Pi^m \lambda^\alpha (ik^1_m) W^{1 \xi}  D_\xi A^2_\alpha:,
\end{align}
\begin{align}
     \oint_{z_2} d z_1 : \frac{1}{2} N^{mn} F^1_{m n}(z_1) : \lambda^\alpha A^2_\alpha(z_2) &= : N^{m n} \lambda^\alpha \big( \frac{1}{2} F^1_{ mn} A^2_\alpha \big) : \nonumber \\ &- : \partial \theta^\beta \lambda^\alpha D_\beta \big( \frac{1}{4} F^1_{pq} (\gamma^{pq})_\alpha^{\ \xi} \big) A^2_\xi  \nonumber \\ &- : \Pi^m \lambda^\alpha \big( ik^1_m \frac{1}{4} (\gamma^{pq})_\alpha^{\ \xi} F^1_{pq} A^2_{\xi } \big) :.
\end{align}
The vertex operator can therefore be written as
\begin{align}
    V^{(12)}_{m^2 = 2} = : \partial \theta^\beta \lambda^\alpha \Bar{B}_{\alpha \beta}: &+ : \Pi^m \lambda^\alpha \Bar{H}_{m \alpha} : + : d_\beta \lambda^\alpha \Bar{C}^{\beta}_{\ \alpha} : + :\frac{1}{2} N^{mn} \lambda^\alpha \Bar{F}_{ m n \alpha } : , \label{answer}
\end{align}
with 
\begin{align}
    \Bar{B}_{\alpha \beta} &= - (\gamma^m W^1)_\beta  (ik^2_m) A^{2}_\alpha - D_\beta W^{1 \xi} D_\xi A^2_\alpha  - D_\beta D_\alpha W^{1 \xi} A^2_\xi  \label{campo_B} \\ 
    \Bar{H}_{m \alpha} &=  A_m^1 A_\alpha^2 \nonumber \\ &- (ik^1_m)\big(A^1_n ik^n_2 A^2_\alpha + W^{1 \xi} D_\xi A_\alpha^2 +  D_\alpha W^{1 \xi} A^2_{\xi} \big), \label{campo_H} \\
    \Bar{C}^{\beta}_{\ \alpha} &= W^{1 \beta} A^2_\alpha,    \label{campo_C} \\
    \Bar{F}_{ m n \alpha } &= F^1_{mn} A^2_\alpha .     \label{campo_F}
\end{align}
It is BRST invariant by construction, as one can see by applying the onshell condition and the descent relation for $U^{(1)}$ and $V^{(2)}$. But one can check how BRST charge acts on each term of \ref{answer},
\begin{align}
     Q  : \Pi^m \lambda^\alpha  A^{1}_m A^{2}_\alpha   :\ = &+ :(\gamma^m \partial \theta)_\alpha \lambda^\alpha \lambda^\beta A^1_m A_\beta^2 : +: \Pi^m \lambda^\alpha \lambda^\beta (D_\alpha A_m^1) A_\beta^2 : \end{align}
     \begin{align} Q   : \Pi^n (i k^1_n) A_m^{1}   \lambda^\alpha \partial^m A^{2}_\alpha : = &+ \ : (\gamma^n \partial \theta)_\alpha \lambda^\alpha \lambda^\beta (i k^1_n)(i k^{2 m}) A^1_{m} A^2_\beta:  \nonumber \\ &+ : \Pi^n \lambda^\beta \lambda^\alpha (ik^1_n) (i k^{2 m}) D_\alpha A^1_m A^2_\beta : \end{align} 
    \begin{align} Q  :   i k_{1m} \Pi^m  W^{1\alpha}   \lambda^\beta D_\alpha  A^{2}_\beta : = &+  : \lambda^\xi (\gamma^m \partial \theta)_\xi \lambda^\beta (ik^1_m) W^{1 \alpha} D_\alpha A^2_\beta :
    \nonumber \\ &+:\Pi^m \lambda^\beta (i k^1_m) \lambda^\xi D_\xi W^{1 \alpha} D_\alpha A^2_\beta: \nonumber \\ &- : \Pi^m \lambda^\beta (i k^1_m) W^{1\alpha} \lambda^\xi D_\xi D_\alpha A^2 : \end{align}
    \begin{align}  Q  :  i k_{1p} \Pi^p   \lambda^\alpha F^{1}_{mn}   (\gamma^{mn})^\beta_{\ \alpha} A^{2}_\beta  :  = &+
 : (\gamma^p\partial \theta)_\alpha \lambda^\alpha \lambda^\xi (i k^1_p) F^1_{mn} (\gamma^{mn})^\beta_{\ \xi} A^2_\beta : \nonumber \\ &+  : \Pi^p \lambda^\xi (ik^1_p) \lambda^\alpha D_\alpha F^1_{mn} (\gamma^{mn} )^\beta_{\ \xi} A^2_\beta : \nonumber \\ &+   : \Pi^p \lambda^\xi (ik^1_p ) F^1_{mn} (\gamma^{mn})^\beta_{\ \xi} \lambda^\alpha D_\alpha A^2_\beta : \end{align}
   \begin{align}  Q  : d_\alpha \lambda^\beta W^{1\alpha}  A^{2}_\beta   : = &- : \Pi_m \lambda^\xi \lambda^\beta \gamma^m_{\xi \alpha } W^{1 \alpha } A^2_\beta : \nonumber \\ &+ :\partial \lambda^\xi \lambda^\beta \gamma^m_{\xi \alpha} (i k^{12}_m) W^{1 \alpha} A^2_\beta  :  \nonumber \\ &- : d_\alpha   \lambda^\xi \lambda^\beta D_\xi W^{1 \alpha } A^2_\beta : \label{QV5} \end{align}
    \begin{align}
         Q  :- \partial \theta^\xi \lambda^\beta D_\xi W^{1\alpha}    D_\alpha A^{2}_\beta : = &- :\partial \lambda^\alpha \lambda^\beta D_\alpha W^{1 \xi} D_\xi A^2_\beta : \nonumber \\ &+: \partial \theta^\alpha \lambda^\beta \lambda^\xi D_\xi \big( D_\alpha W^{1 \gamma} D_\gamma A^2_\beta \big):  \end{align}
     \begin{align} Q :- \partial \theta^\alpha \lambda^\beta D_\alpha A_m^{(1)}   \partial^m A^{(2)}_\beta : = &- : \partial \lambda^\alpha \lambda^\beta (\gamma_m W^1)_\alpha (ik^m_2) A_\beta^2 :  \nonumber \\ &- : \partial \lambda^\alpha \lambda^\beta  A^1_\alpha A^2_\beta :  \nonumber \\ &+:\partial \theta^\alpha \lambda^\xi \lambda^\beta  \gamma_{m \alpha \gamma} D_\xi W^{1 \gamma} (ik^m_2) A_\beta^2 : \nonumber \\ &+ : \partial \theta^\alpha \lambda^\xi \lambda^\beta D_\xi A_\alpha^1 A_\beta^2 :  \end{align}
    \begin{align} Q  :  \partial \theta^\alpha \lambda^\beta
A^{1}_\alpha   A^{2}_\beta   : = &+ \ : \partial \lambda^\alpha \lambda^\beta A^1_\alpha A^2_\beta : - : \partial \theta^\alpha \lambda^\beta \lambda^\xi D_\xi A^1_\alpha A^2_\beta :  \end{align}
\begin{align}  Q  : \partial \theta^\alpha D_\alpha  F^{1}_{mn}   (\gamma^{mn}\lambda)^\beta A^{2}_\beta  :  = &-  : \partial \lambda^\alpha \lambda^\xi D_\alpha F^1_{mn} (\gamma^{mn})^\beta_{\ \xi} A^2_\beta :   \nonumber \\ &- : \partial \theta^\alpha \lambda^\delta \lambda^\xi D_\xi (D_\alpha F^1_{mn}) (\gamma^{mn})^\beta_{\ \delta} A^2_\beta : \nonumber \\ &+ :\partial \theta^\alpha \lambda^\xi \lambda^\gamma D_\alpha F^1_{mn} (\gamma^{mn})^\beta_{\ \xi} D_\gamma A^2_\beta : \end{align} \begin{align}  Q  : \frac{1}{2} N^{mn} \lambda^\beta  F^{1}_{mn}  A^{2}_\beta  : = &- \frac{1}{4} : (\gamma^{mn})^\alpha_{\ \xi} \partial \lambda^\xi \lambda^\beta D_\alpha (F^1_{mn} A^2_\beta) : \nonumber \\ &-\frac{1}{4}:(\gamma^{mn})^\alpha_{\ \xi} d_\alpha \lambda^\xi \lambda^\beta F^1_{mn} A^2_\beta : \nonumber \\ &+ \frac{1}{2} : N^{mn} \lambda^\beta \lambda^\alpha D_\alpha F^1_{mn} A^2_\beta :. \label{QV8} \end{align}

Collecting each ghost number 2 component proportional to $\partial \theta^\xi \lambda^\alpha \lambda^\beta$, $\Pi^m \lambda^\alpha \lambda^\beta$, $ \partial \lambda^\alpha \lambda^\beta$, $d_\xi \lambda^\alpha \lambda^\beta $ and $N^{mn} \lambda^\alpha \lambda^\beta$, one can see that the BRST variation of $V^{(12)}_{m^2 = 2}$ vanishes. For example, the terms proportional to $d_\xi \lambda^\alpha \lambda^\beta$ in \ref{QV5} and \ref{QV8} cancel each other. Using the equation of motion $ik^1_m (\gamma^m W^1 )_\alpha = 0 $ and the pure spinor identity $ (\lambda \gamma^n )_\alpha (\lambda \gamma_n )_\beta = 0$, one can show that the following constraint \cite{Berkovits:2002qx} \begin{align}
 : N^{mn} \lambda^\beta \lambda^\alpha : (\gamma_{m})_{\alpha \gamma} &= \frac{1}{2} :J \lambda^\beta \lambda^\alpha: \gamma^n_{\alpha \gamma}  + \frac{5}{2} \lambda^\beta \partial \lambda^\alpha \gamma^n_{\alpha \gamma} + \frac{1}{2} \lambda^\delta \partial \lambda^\alpha (\gamma^{sn})_{\delta}^{\ \beta} (\gamma_{s})_{\alpha \gamma},
\end{align} implies that the last term of \ref{QV8} can be written as
\begin{align}
    :\frac{1}{2} N^{mn} \lambda^\alpha \lambda^\beta D_\alpha F^1_{mn} A^2_{\beta } : \ = \  : \frac{1}{4} \partial \lambda^\alpha \lambda^\delta (\gamma^{ns})_{\delta}^{\ \beta} D_\alpha F^1_{n s} A^2_\beta :,
\end{align} and therefore cancels all other terms proportional to $\partial \lambda^\alpha \lambda^\beta$.\footnote{I would like to thank Carlos Mafra for correcting an error in an earlier version of this computation.}

Physical information of \ref{answer} is obtained through a gauge fixing procedure wherein the massive vertex operator superfields are related to the spin-2 massive supermultiplet in 10 dimensions. This multiplet comprises a traceless symmetric tensor denoted as $g_{mn}$, a three form $b_{mnp}$ and a spin-$\frac{3}{2}$ field $\psi_{m \alpha}$, all satisfying  \begin{equation}
     (\gamma^m)^{\beta \alpha} \psi_{m \alpha} = 0, \ \ \partial^m g_{mn} = 0, \ \  \partial^m \psi_{m \alpha} = 0, \ \ \partial^m b_{mnp} = 0.
\end{equation}

\section{Gauge transformations} \label{sec3}

In this section, the operator vertex \ref{answer} will be gauge fixed following the procedure of \cite{Berkovits:2002qx} to a gauge where  \begin{align}
     B_{\alpha \beta} &= \gamma_{\alpha \beta}^{mnp} B_{mnp},  &\partial^m B_{mnp} = 0, \label{gauge_fixing_procedure1} \\
     \gamma^{m \alpha \beta} H_{m \beta} &= 0,  &\partial^m H_{m \alpha} = 0,   \label{gauge_fixing_procedure2}\\
      C^{\beta}_{\ \alpha} &= (\gamma^{mnpq})^{\beta}_{\ \alpha } C_{mnpq} ,   &\gamma^{m \alpha \beta} F_{\alpha m n } = 0. \label{gauge_fixing_procedure3}
 \end{align} 
 In this gauge, one can check that the $\theta = 0$ 
 components of the superfields $B_{mnp}$, $G_{mn} \equiv 
 D_{\alpha} \gamma^{\alpha \beta}_{(m} H_{n) \beta}$ and $-
 \frac{1}{72}H_{m\alpha}$ are $b_{mnp},g_{mn},\psi_{m \alpha}$ 
 respectively \cite{Berkovits:2002qx}.

The operator vertex \ref{answer} is gauge invariant by  
\begin{equation}
    \label{gaugetr}
    V^{(12)}_{m^2 = 2} \longrightarrow V^{(12)}_{m^2 = 2} + Q \Omega,
\end{equation}
where  \begin{align} \Omega = &+ :\partial \theta^\alpha \Omega_{1 \alpha}  : + :d_\alpha \Omega^\alpha_2  :  + : \Pi^m \Omega_{3 m} : + : J \Omega_4 : + : N^{mn} \Omega_{5 mn} :.
\end{align}
Using the OPE's \ref{OPES}, one finds
\begin{align}
    Q \Omega &=  : \partial \lambda^\alpha \bigg( \Omega_{1 \alpha} + \gamma^m_{\alpha \xi} \partial_m \Omega_2^\xi -  D_\alpha \Omega_4 - \frac{1}{2} (\gamma^{mn})^\beta_{\ \alpha} D_\beta \Omega_{5 mn} \bigg) : \nonumber \\ 
    &+ : \partial \theta^\beta \lambda^\alpha \bigg(  - D_\alpha \Omega_{1 \beta} + \gamma^{m}_{\alpha \beta} \Omega_{3 m} \bigg): \nonumber \\
    &+ : \Pi^m \lambda^\alpha \bigg( - \gamma_{m \alpha \xi} \Omega^\xi_2 + D_\alpha \Omega_{3 m}  \bigg): \nonumber \\
    &+ : d_\beta \lambda^\alpha \bigg(  - D_\alpha \Omega_2^\beta - \delta^\beta_\alpha \Omega_4 - \frac{1}{2}(\gamma^{mn})^\beta_{\ \alpha} \Omega_{5 mn} \bigg): \nonumber \\
    &+ : N^{mn} \lambda^\alpha \bigg( D_\alpha \Omega_{5 mn}  \bigg): \nonumber \\
    &+ : J \lambda^\alpha \bigg( D_\alpha \Omega_4 \bigg) :, 
\end{align} 
so the vertex operator superfields have the following variations
 \begin{align}
     \delta \Bar{B}_{\alpha \beta} &= - D_\alpha \Omega_{1 \beta} + \gamma^{m}_{\alpha \beta} \Omega_{3 m},     \label{Bgaugetr}\\ 
     \delta \Bar{H}_{m \alpha} &= - \gamma_{m \alpha \xi} \Omega^\xi_2 + D_\alpha \Omega_{3 m}, \label{Hgaugetr} \\
     \delta \Bar{C}^{\beta}_{\ \alpha} &= - D_\alpha \Omega_2^\beta - \delta^\beta_\alpha \Omega_4 - \frac{1}{2}(\gamma^{mn})^\beta_{\ \alpha} \Omega_{5 mn}, \label{Cgaugetr} \\
     \delta \Bar{F}_{ m n \alpha } &= D_\alpha \Omega_{5 mn}.    \label{Fgaugetr}
 \end{align}

There are additional terms proportional to $\partial \lambda^\alpha$ and $J \lambda^\alpha$  coming from the gauge transformation \ref{gaugetr},
\begin{align}
    \bar{G}_\alpha &\equiv   \Omega_{1 \alpha} + \gamma^m_{\alpha \xi} \partial_m \Omega_2^\xi -  D_\alpha \Omega_4 - \frac{1}{2} (\gamma^{mn})^\beta_{\ \alpha} D_\beta \Omega_{5 mn}, \\
    \bar{E}_\alpha &\equiv D_\alpha \Omega_4,
\end{align}
and the following constraint \cite{Berkovits:2002qx}
\begin{equation}
    : N^{mn} \lambda^\alpha \gamma_{m \alpha \beta}: - \frac{1}{2} : J \lambda^\alpha  \gamma^n_{\alpha \beta}: - 2 \partial \lambda^\alpha \gamma^n_{\alpha \beta} = 0
\end{equation}
implies that \ref{answer} is invariant under the field redefinition
\begin{align}
    \delta_{\Lambda} \bar{G}_\alpha &= - 4 \gamma^n_{\alpha \xi} \Lambda^\xi_{n}, \label{newgaugeG}\\
    \delta_{\Lambda} F_{\alpha m n} &= \gamma_{m \alpha \xi} \Lambda^\xi_{n} - \gamma_{n \alpha \xi} \Lambda^\xi_{m}, \label{newgaugeF}  \\
    \delta_{\Lambda} \bar{E}_\alpha &= - \gamma^n_{\alpha \xi} \Lambda^\xi_n. \label{newgaugeE}
\end{align}

Finally, after the gauge-fixing procedure \ref{gauge_fixing_procedure1}, \ref{gauge_fixing_procedure2}, \ref{gauge_fixing_procedure3}, all vertex operator superfields will be expressed in terms of d=10 Yang-Mills superfields and will satisfy the equations:
\begin{align}
    H_{m \alpha} &= \frac{3}{7} (\gamma^{st})_{\alpha}^{\beta}D_\beta B_{mst}, \label{relationHandB} \\ C^{\alpha}_{\beta} &= \frac{1}{4} (\gamma^{ mnpq})^{\alpha}_{\ \beta} \partial_{m}B_{npq},  \label{relationCandB}  \\  F_{mn \alpha} &= \frac{1}{16} \big( 6 \mathcal{H}_{mn \alpha} -  (\gamma_{p [m})_{\alpha}^{\ \beta} \mathcal{H}_{n] p \beta} \big), \label{relationFandH} \\ E_\alpha &= 0, \label{vanishingofE} \\
    G_\alpha  &= 0 \label{vanishingofG}
\end{align}
where $\mathcal{H}_{mn\alpha} \equiv \partial_{[m} H_{n] \alpha}$. The above equations and $(\partial^m \partial_m - 2)V^{(12)}_{m^2=2} = 0$ imply that \ref{answer} describes a massive spin-two multiplet with $(mass)^2 = 2$ \cite{Berkovits:2002qx}.

\subsection{Fixing B and H}
In this subsection, the $42$ degrees of freedom of $\Omega_{1\beta}, \Omega_{2}^{\xi}$, $  \Omega_{3 m }$ will be used to impose the following constraints on $\Bar{B}_{\alpha \beta}$ and $\Bar{H}_{m \beta}$  \begin{align}
 \label{algebraic_conditions_B}
     B_{\alpha \beta} &= \gamma^{m n p}_{\alpha \beta} B_{m n p}, \\  \label{algebraic_conditions_H}  H_{m \beta} \gamma^{m \beta \alpha} &= 0.
 \end{align}


Using Super Yang-Mills equations of motion \ref{sym_eom:a}, \ref{sym_eom:b}, \ref{sym_eom:c}, and Fierz decomposition \ref{fierzbb} the bi-spinor \ref{campo_B} can be written as
\begin{equation}
    \Bar{B}_{\alpha \beta } \equiv  \gamma^{m_1}_{\alpha \beta}  \Bar{B}_{m_1}  +   \gamma^{m_1 m_2 m_3}_{\alpha \beta}  \Bar{B}_{m_1 m_2 m_3}  + \gamma^{m_1 m_2 m_3 m_4 m_5}_{\alpha \beta}   \Bar{B}_{m_1 m_2 m_3 m_4 m_5}, \label{fierzB} 
\end{equation}
where
\begin{align}
    B_{m_1} &= -\frac{1}{2} W^1 \gamma_{m_1} W^2 - F^1_{m_1 m} A^2_m - (ik^1_{m_1}) W^{1\xi} A^2_{\xi}  \\
    &+ \frac{\gamma^{\alpha \beta}_{m_1}}{16}   D_\alpha \bigg( (\gamma^m W^1)_\beta A^2_m + D_\beta W^{1 \xi} A^2_\xi \bigg) , \nonumber
\end{align}
\begin{align}
     B_{m_1 m_2 m_3} &= \frac{1}{24}  W^1 \gamma_{m_1 m_2 m_3} W^2 
      \\ &+ \frac{\gamma^{\alpha \beta}_{m_1 m_2 m_3}}{96}   D_\alpha \bigg( (\gamma^m W^1)_\beta A^2_m + D_\beta W^{1 \xi} A^2_\xi \bigg),  \nonumber 
     \end{align} \begin{equation}
      B_{m_1 m_2 m_3 m_4 m_5} = \frac{\gamma^{\alpha \beta}_{m_1 m_2 m_3 m_4 m_5}}{3840}  D_\alpha \bigg( (\gamma^m W^1)_\beta A^2_m + D_\beta W^{1 \xi} A^2_\xi \bigg). 
\end{equation}
To obtain the algebraic condition \ref{algebraic_conditions_B}, one can choose
 \begin{align}
     \Omega'_{1 \gamma} &= (\gamma^m W^1)_\gamma A^2_m + D_\gamma W^{1 \xi} A^2_\xi, \label{old_variations_1} \\
  \Omega'_{3 m} &= \frac{1}{2} (W^1 \gamma_m W^2) + F^1_{m n} A^{2 n } + (ik^1_m) W^{1 \xi} A^2_\xi \label{old_variations_3},
\end{align}
and \ref{algebraic_conditions_H} is therefore implied by, 
\begin{gather}
     \Omega^{' \beta}_2 = \frac{1}{10} \big[ - 7 D_\xi W^{1 \beta} W^{2 \xi} - 10 (i k^1_n) W^{1\beta} A^{2 n} + 3 W^{1 \xi } D_\xi W^{2 \beta} \big].\label{old_variation_2}
\end{gather}

 In this gauge, $B'_{mnp} = \frac{1}{96} \gamma^{\alpha \beta}_{mnp} (\Bar{B}_{\alpha \beta} + \delta \bar{B}_{\alpha \beta})$ is
\begin{equation}
    B'_{m n p } = \frac{1}{24} W^1 \gamma_{mnp} W^2, \label{Bingauge}
\end{equation}

and $H'_{m \alpha} = \Bar{H}_{m \alpha} + \delta \bar{H}_{m \alpha} $ is
\begin{gather}
    H'_{m \alpha} = \bigg( - \frac{8}{20} \gamma^{p}_{\alpha \xi} \delta^{q}_{m} - \frac{1}{20} \gamma^{ m p q}_{\alpha \xi } \bigg) \bigg( F^1_{p q} W^{2 \xi} + F^2_{ p q} W^{1 \xi } \bigg), \label{oldH_with_SYM}
\end{gather}
which is traceless, as one can verify by using $\gamma^{m \beta \alpha } (\gamma_{p q m })_{\alpha \xi} = 8 (\gamma_{p q})^\beta_{\ \xi} $. 

To understand the relation between \ref{Bingauge} and \ref{oldH_with_SYM}, one can define the tensor 
\begin{equation}
    H^{B'}_{m \alpha } := (\gamma^{n p})_{\alpha}^{\ \beta} D_\beta B'_{m n p}. \label{oldHfromBtilde}
\end{equation} It can be expressed from \ref{Bingauge} as
\begin{gather}
   H^{B'}_{m \alpha } 
  = \bigg( - \frac{10}{12} \gamma^p_{\alpha \xi }\delta^{q}_{m} - \frac{2}{12} \gamma^{m p q}_{\alpha \xi} \bigg) \bigg( F^1_{p q} W^{2 \xi} + F^2_{ p q} W^{1 \xi } \bigg),
\end{gather}
and has a non-vanishing trace
\begin{equation}
   F^\beta \equiv \gamma^{m \beta \alpha} H^{B'}_{m \alpha} = 2 D_\xi \big( W^{1 [\beta} W^{2 \xi ]} \big). \label{traceparttildeH}
\end{equation}

It will be useful to note that the traceless part $(H^{B'})^{(0)}_{m \alpha} \equiv H^{B'}_{m \alpha } - (\gamma_{m})_{ \alpha \xi} \bigg( \frac{1}{10}F^{\xi} \bigg)$ of \ref{oldHfromBtilde} satisfies the relation
\begin{equation}
     H'_{s \alpha } = \frac{3}{7} (H^{B'})^{(0)}_{s \alpha}. \label{almost_BandHrelation}
\end{equation}
Nevertheless, the expression \ref{Bingauge} for $B'_{mnp}$ does not satisfy the transversality condition. This is a necessary condition to remove the extra degrees of freedom at the zeroth order in $\theta$
expansion of $B'_{mnp}$ and $H'_{m\alpha}$ \cite{Chakrabarti:2017vld}.

\subsection{Additional gauge-fixing}

In this subsection, it will be shown that $\partial^m B_{mnp} = 0$, when $\Omega_{1 \beta}$ is written as \begin{equation}
    \Omega_{1 \beta} =  \Omega'_{1 \beta} + D_\beta \Lambda. \label{omega1completo}
\end{equation}
In this gauge, $B_{\alpha \beta}$ and $H_{m\alpha}$ are related as \ref{relationHandB}.
  
The additional contribution $\Omega^{(1)}_{1 \beta} = D_\beta \Lambda$ does not change the five-form part of $B_{\alpha \beta}$ because of the identity $\gamma^{\alpha \beta}_{mnpqr} D_{\alpha} D_{\beta} = 0$. So the previous subsection gauge fixing leaves gauge invariances parameterized by $\Omega^{(1)}_{1 \beta}$. After this additional gauge-fixing, the resulting $B_{mnp}$ is \begin{equation}
    B_{mnp} = \frac{1}{24} W^1 \gamma_{mnp} W^2 - \frac{1}{96} \gamma^{\alpha \beta}_{mnp} D_\alpha  \Omega^{(1)}_{1 \beta}. \label{New_Bfield}
\end{equation}

To obtain $\Lambda$ in terms of SYM superfields, $H^{B}_{m \alpha} := (\gamma^{n p})_{\alpha}^{\ \beta} D_\beta B_{m n p}$ will be required to satisfy $\gamma^{m \alpha \beta} H^{B}_{m \alpha} = 0$. Indeed, if $H^{B}_{m \alpha}$ is assumed to be traceless, \ref{traceparttildeH} implies that
\begin{gather}
(\gamma^{ m s t})^{\beta \xi} D_\xi \bigg( -\frac{1}{96}\gamma^{\delta \alpha}_{mst} D_\delta  \Omega^{(1)}_{1 \alpha} \bigg) = - 2 D_\xi \bigg( W^{1 [ \beta }W^{2 \xi] } \bigg).  \label{conditionHtraceless}
\end{gather} Hitting both sides of \ref{conditionHtraceless} with $D_\beta$, one finds that
\begin{equation}
   \frac{1}{96} (D \gamma^{mnp} D)(D \gamma_{mnp} D) \Lambda =  2 D_\beta  D_\xi \bigg( W^{1 [ \beta }W^{2 \xi] } \bigg).
\end{equation} But $(D \gamma^{mnp} D)(D \gamma_{mnp} D)  =  96 \cdot 48$ at the first massive level, 
then $\Lambda$ is given by 
\begin{gather}
   \Lambda = -\frac{1}{6}  F^1_{mn} F^2_{mn}, \label{Lambda_asFF} \end{gather}
and the additional gauge fixing $\Omega^{(1)}_{1 \beta}$ is
\begin{gather}
    \Omega^{(1)}_{1 \beta}  = - \frac{1}{3} \bigg[ ik^1_m (\gamma_n W^1)_\beta F^2_{mn} + (1 \leftrightarrow 2 )  \bigg]. \label{additional_omega1}
\end{gather} 


In the gauge $\gamma_m^{\alpha \beta} B_{\alpha \beta} = 0$, $\gamma^{m \alpha \beta} H_{m \alpha} = 0$, one has 
\begin{align}
\Omega_{3m } &\equiv \Omega'_{3m} + \Omega^{(1)}_{3m} = \frac{1}{2} (W^1 \gamma_m W^2) + F^1_{m n} A^{2 n } + (ik^1_m) W^{1 \xi} A^2_\xi + \frac{1}{2} \partial_m \Lambda, \label{Omega_3} 
\end{align} and
\begin{align}
   \Omega_{2}^{\beta} &\equiv \Omega^{' \beta}_{2} + \Omega^{(1) \beta}_{2} =  - \partial^m \big( W^{1 \beta } A^{2}_m \big) - \frac{2}{3} D_\alpha W^{1 \beta} W^{2 \alpha} + \frac{1}{3} W^{1 \alpha} D_\alpha W^{2 \beta}, \label{Omega_2}
\end{align}
and $B_{mnp}$ is transverse to $k_1 + k_2$ because of
\begin{equation}
      \partial^m \bigg(- \frac{1}{96} \gamma^{\alpha \beta}_{mnp} D_\alpha D_\beta \Lambda \bigg) =  -\partial^m \big( \frac{1}{24}  W^1 \gamma_{mnp} W^2 \big).
\end{equation}

To demonstrate \ref{relationHandB}, one can write
\begin{align}
   H_{m \alpha} &\equiv H'_{m \alpha} + \delta H_{m \alpha} \label{newFullH} \\
   H^{B}_{m \alpha} &\equiv  H^{B'}_{m \alpha} + \delta H^{B'}_{m \alpha} , \label{new_HfromB}
\end{align}
where $\delta H_{m \alpha} = - (\gamma_m)_{\alpha \xi} \Omega_{2}^{ (1)\xi} + D_\alpha \Omega^{(1)}_{3m}$ is the variation of \ref{oldH_with_SYM}, \begin{align}
     \delta H_{m \alpha} &= - \frac{1}{3\cdot 10} (\gamma_m)_{\alpha \xi} \big( D_\beta W^{1 \xi} W^{2 \beta} + W^{1 \beta} D_\beta W^{2 \xi}  \big)  + D_\alpha \big( \frac{1}{2} \partial_m \Lambda \big),  \label{variation_fullH}
\end{align} and $\delta H^{B'}_{m \alpha}$ is the variation of \ref{oldHfromBtilde}
\begin{equation}
    \delta H^{B'}_{m \alpha} = (\gamma^{st})_{\alpha}^{\beta} D_\beta \big( -\frac{1}{96} \gamma^{\gamma \delta}_{mst} D_\gamma D_\delta \Lambda \big), \label{variation_HfromB}
\end{equation}
which is implied by \ref{oldHfromBtilde}, \ref{New_Bfield} and \ref{new_HfromB}. Using \ref{almost_BandHrelation}, one can write a statement equivalent to \ref{relationHandB},
\begin{equation}
    \delta H_{m \alpha} = \frac{3}{7} \big( \delta H^{B'}_{m \alpha}  + \frac{1}{10}  (\gamma_m)_{\alpha \beta} F^\beta \big),\label{equivalent_relationHeB}
\end{equation}
with $ F^{\beta}$ defined in \ref{traceparttildeH}. One finds from the identity
\begin{equation}
    (\gamma^{st})^{\ \beta}_{\alpha} (\gamma_{mst})^{\gamma \delta} D_{\beta} D_{\gamma} D_{\delta } = - 72 \partial_m D_{\alpha } + 40 (\gamma_{mt})^{\ \beta}_{\alpha} \partial^t,
\end{equation}
that the variation \ref{variation_HfromB} is
\begin{gather}
      \delta H^{B'}_{m \alpha} = \frac{7}{6} \partial_m D_\alpha \Lambda - \frac{5}{36} (\gamma_m)_{\alpha \beta} F^{\beta},
\end{gather}
and \ref{equivalent_relationHeB} is therefore satisfied,
\begin{equation}
    \delta H^{B'}_{m \alpha} + \frac{1}{10} (\gamma_m)_{\alpha \beta} F^\beta = \frac{7}{3} \bigg[ \frac{1}{2} \partial_m D_\alpha \Lambda - \frac{1}{60} (\gamma_m)_{\alpha \beta} F^\beta \bigg]. 
\end{equation}
 So it has been proven that in the gauge \ref{algebraic_conditions_B}, \ref{algebraic_conditions_H} and $\partial^m B_{mnp} = 0$, the equation \ref{relationHandB} is satisfied. 
 
 In this gauge, the superfield $H_{m \alpha}$ is \begin{align}
    H_{m \alpha} &=  \bigg[  \partial^m \partial^n \frac{1}{6} \gamma^p_{\alpha \beta} \delta^{q}_n - \frac{10}{24} \gamma^p_{\alpha \beta} \delta^{q}_m - \frac{1}{24} (\gamma_{m p q})_{\alpha \beta} \bigg] \big( F^{1 pq} W^{2 \beta} + F^{2 pq} W^{1 \beta} \big),  \label{FinalH}
\end{align}


\subsection{Fixing C}

In this subsection, the $46$ degrees of freedom of $\Omega_{4}$ and $\Omega_{5 mn}$ will be used to impose the algebraic constraint 
\begin{equation}    
    \label{algebraic_conditions_C}
    C^{\beta}_{\ \alpha} = (\gamma^{ m n p q})^\beta_{\ \alpha} C_{m n p q}.
\end{equation}
From the Fierz decomposition \ref{fierzbc}, 
one finds
\begin{align}
     \Omega_4 &= - \frac{1}{24} F^1_{mn} F^{2 mn}, \label{finalOmega4}\\
      \Omega_{5 p q} &= \frac{1}{16} (\gamma_{pq})^{\ \alpha}_{\beta} \big[ W^{1 \beta} A^2_{\alpha} - D_\alpha \Omega^\beta_2 \big]. \label{omega_sem_expandir} 
\end{align}
Using \ref{Omega_2}, $\Omega_{5 mn}$ is \begin{align}
    \Omega_{5mn} &= \frac{1}{2} F^1_{mn} (ik^1 \cdot A^2) + \frac{1}{4} \partial_{[m} W^1 \gamma_{n]} W^2 - \frac{1}{8} \partial^r W^1 \gamma_{mnr} W^2 + \frac{1}{4} F^1_{p [m} F^2_{n] p}. \label{omega5final} 
\end{align}
The $\gamma^{(4)}$ component of $C^{\beta}_{\ \alpha}$ is
\begin{align}
C^{mnpq} &= \frac{(\gamma^{m n p q })^{\ \alpha }_{\beta}}{384}    \bigg( \Bar{C}^\beta_{\ \alpha} - D_\alpha \Omega^\beta_{2} - \delta^{\beta}_{\ \alpha} \Omega_4 - \frac{(\gamma^{pq})^\beta_{\ \alpha}}{2}  \Omega_{5 pq} \bigg), 
\end{align} one therefore obtains from \ref{Omega_2}, \ref{finalOmega4}, \ref{omega5final} that
 \begin{gather} C^{mnpq} =  \frac{1}{96 \cdot 12}  F^1_{[mn} F^2_{pq]}+ \frac{1}{96 \cdot 36}  \partial_{[m} W^1 \gamma_{npq]} W^2 . \label{C_final}
\end{gather}

Finally, the equation
\begin{align}
      - \frac{1}{96} \partial_{[ m} \gamma^{\alpha \beta}_{npq]} D_\alpha D_\beta \Lambda &=  \frac{1}{12} F^1_{[mn} F^2_{pq]} - \frac{1}{72} \partial_{[m} W^1 \gamma_{npq]} W^2 \label{partial_shiftnewB} 
\end{align}
implies that \ref{New_Bfield} and \ref{C_final} are related as
\begin{equation}
    C_{mnpq} = \frac{1}{96} \partial_{[m} B_{npq]}.
\end{equation}

\subsection{Fixing F}
In this subsection, the gauge invariance  \ref{newgaugeF} with
\begin{equation}
    \Lambda^{\beta}_n = (\gamma_{n})^{\beta \alpha} \big( \frac{1}{10} (\gamma^n)_{\alpha \xi} \Lambda^\xi_n \big) + \Lambda^{(0)\beta}_n \label{gauge_parameter_Lambda}
\end{equation} will be used to impose the following algebraic constraint
\begin{align}
    \gamma^{m \beta \alpha } \big[  \frac{1}{2} \Bar{F}_{m n \alpha} +  \delta \Bar{F}_{m n \alpha}  +\delta_{\Lambda} \Bar{F}_{m n \alpha}  \big] &= 0, \label{condition_on_F}\\
    \bar{E}_\alpha + \delta_\Lambda E_\alpha &= 0. \label{conditiononE}
\end{align}

To obtain \ref{conditiononE}, the trace part of \ref{gauge_parameter_Lambda} should be  \begin{equation}
     \gamma^n_{\alpha \beta} \Lambda^\beta_n = D_\alpha \Omega_4,
\end{equation}
so the constraints \ref{condition_on_F}, \ref{conditiononE} imply
\begin{equation}
        \Lambda^{\beta}_{n} = -\frac{1}{8} \gamma^{m \beta \alpha }  \big[ \frac{1}{2} F^1_{mn} A^2_\alpha + D_\alpha \Omega_{5 mn} \big] -\frac{1}{8} \gamma^{\beta \alpha}_n D_\alpha \Omega_4.\label{condition_on_F+E}
\end{equation}
In this gauge, $\frac{1}{2} F_{mn \alpha}$ can be written as 
\begin{align}
    \frac{1}{2} F_{mn \alpha} &= \frac{6}{8} \bigg( \frac{1}{2} F^1_{mn} A^2_{\alpha } + D_\alpha \Omega_{5 mn} \bigg) - \frac{1}{16} (\gamma_{mn})_\alpha^{\ \beta}  D_\beta \Lambda - \frac{1}{8} (\gamma_{p [m})_\alpha^{\ \beta} \big( \frac{1}{2} F^1_{n]p} A^2_{\beta } + D_\beta \Omega_{5 n]p} \big).  
\end{align} Using the equation $\gamma_{p[m} \gamma_{n]p} = 16 \gamma_{mn}$, one obtains
\begin{equation}
    F_{mn \alpha} = \frac{1}{8} \big( 6 \mathcal{F}_{m n \alpha} - (\gamma_{p [m}    \mathcal{F}_{n] p} )_{\alpha} \big), \label{equationforFmy}
\end{equation}
where \begin{equation}
    \mathcal{F}_{m n \alpha} =  F^1_{mn} A^2_{\alpha } + 2 D_\alpha \Omega_{5 mn} + \frac{1}{10} (\gamma_{mn})_\alpha^{\ \beta}  D_\beta \Lambda. \label{finalexpressionforF}
\end{equation}

To show the relation \ref{relationFandH}, one can add $\mathcal{F}^{(0)}_{mn \alpha}$ to $\mathcal{F}_{mn\alpha}$, such that
\begin{equation}
    6 \mathcal{F}^{(0)}_{mn \alpha} = \gamma_{p [m} \mathcal{F}^{(0)}_{n]p \alpha}. \label{conditiontoredefineF}
\end{equation}
So one can define the following tensors
\begin{align}
    \mathcal{A}^{W^1}_{mn \alpha} &= \partial_{p} (\gamma_{mnp} W^1 )_\alpha F^2_{pq} , \  &\mathcal{A}^{W^2}_{mn \alpha} &= \partial_{p} (\gamma_{mnp} W^2 )_\alpha F^1_{pq} , \nonumber \\
    \mathcal{B}^{W^1}_{mn \alpha} &= \partial_r (\gamma_{[m} W^1 )_\alpha F^2_{n] r},   &\mathcal{B}^{W^2}_{mn \alpha} &= \partial_r (\gamma_{[m} W^2 )_\alpha F^1_{n] r}  \nonumber \\
    \mathcal{M}^{(k^iW^1)}_{mn \alpha}& = ik^i_{[m} (\gamma_{n] pq} W^1)_\alpha F^2_{pq},  \  &\mathcal{M}^{(k^iW^2)}_{mn \alpha} &= ik^i_{[m} (\gamma_{n] pq} W^2)_\alpha F^1_{pq}, \nonumber \\
    \mathcal{N}^{(k^iW^1)}_{mn \alpha} &=  ik^i_{[m} F^2_{n] r} (\gamma^r W^1)_\alpha, \  &\mathcal{N}^{(k^iW^2)}_{mn \alpha} &=  ik^i_{[m} F^1_{n] r} (\gamma^r W^2)_\alpha,
\end{align}
whose combinations \begin{align}
    \mathcal{R}^{W^i}_{mn \alpha  } &= \mathcal{A}^{W^i}_{mn\alpha} + 4 \mathcal{B}^{W^i}_{mn\alpha}, \\
    \mathcal{S}^{W^{i}}_{mn\alpha} &= 2 \mathcal{N}^{(k^iW^i)}_{mn \alpha} + \mathcal{M}^{(k^iW^i)}_{mn \alpha} - 4 \mathcal{B}^{W^i}_{mn \alpha}, \\
    \mathcal{T}^{W^{1}}_{mn\alpha} &= 2 \mathcal{N}^{(k^2W^1)}_{mn \alpha} + \mathcal{M}^{(k^2W^1)}_{mn \alpha}, \\
    \mathcal{T}^{W^{2}}_{mn\alpha} &= 2 \mathcal{N}^{(k^1W^2)}_{mn \alpha} + \mathcal{M}^{(k^1W^2)}_{mn \alpha}, 
\end{align}
satisfy the relation \ref{conditiontoredefineF}. Expanding \ref{finalexpressionforF}, it is straightforward to check that 
\begin{equation}
   \frac{1}{2} \mathcal{H}_{mn\alpha}  = \mathcal{F}_{mn\alpha} + \mathcal{F}^{(0)}_{mn\alpha},
\end{equation}
where
\begin{equation}
    \mathcal{F}^{(0)}_{mn\alpha} = \frac{1}{30}  \mathcal{R}^{W^{1}}_{mn\alpha} + \frac{1}{30}  \mathcal{R}^{W^{2}}_{mn\alpha} + \frac{1}{6} \mathcal{S}^{W^{1}}_{mn\alpha} - \frac{1}{12} \mathcal{S}^{W^{2}}_{mn\alpha} + \frac{1}{24} \mathcal{T}^{W^{1}}_{mn\alpha} - \frac{5}{24} \mathcal{T}^{W^{2}}_{mn\alpha}, \nonumber
\end{equation}
thus \ref{relationFandH} holds. 

The gauge parameter $\Lambda^\beta_n$ degrees of freedom are sufficient to enforce both conditions \ref{condition_on_F} and \ref{conditiononE}. Indeed, the following spinor
\begin{equation}
    \tilde{\Lambda}^{(0)\beta}_n = -\frac{1}{8} \gamma_{m}^{\beta \alpha } \big( \frac{1}{2} F^1_{mn} A^2_{\alpha} + D_\alpha   \Omega_{5 mn }   \big) - \gamma_n^{\beta \alpha } \big( \frac{9}{160} D_\alpha \Lambda \big),
\end{equation}
should be exactly the traceless part of \ref{gauge_parameter_Lambda}, as one can see by subtracting $(\gamma_{n})^{ \beta \xi} \bigg( \frac{1}{10} \gamma^{m}_{\xi \alpha} \Lambda^{\alpha}_{m} \bigg)$ from \ref{condition_on_F+E}. The gamma matrix expression \ref{gammaid8} and super Yang-Mills equations of motion implies that
\begin{equation}
   \gamma^{n}_{\xi \beta}  \tilde{\Lambda}^{(0)\beta}_n = - \frac{3}{8} D_\xi \Lambda - \frac{3}{16} \gamma^n_{\xi \beta} \big(  W^{1 \beta } A^2_{n} + \partial_n \Omega^\beta_{2} \big). \label{check_Lambda_trace}
\end{equation}
After expressing \ref{check_Lambda_trace} in terms of SYM superfields, one obtains
\begin{equation}
    \gamma^{n}_{\xi \beta}  \tilde{\Lambda}^{(0)\beta}_n = \frac{\partial^m}{32}  \bigg( 2  (\gamma_n W^2)_\xi F^1_{mn} - (\gamma_n W^1)_\xi F^2_{mn}  \bigg) - \frac{\partial_n}{2\cdot 32}  \bigg(  2  (\gamma^{pqn } W^2)_\xi F^1_{pq}  - (\gamma^{pqn} W^1)_\xi F^2_{pq}   \bigg), 
\end{equation}
which vanishes by expanding the second term of the right-hand side with equation \ref{formid0}. 

Finally, it will be shown that $\Bar{G}_\alpha + \delta_\Lambda G_\alpha =0$. Using \ref{newgaugeG}, $G_{\alpha}$ can be written as \begin{equation}
    \bar{G}_\alpha + \delta_\Lambda G_\alpha = \Omega'_{1 \alpha } + \gamma^m_{\alpha \beta} \partial_m \Omega^{\beta}_2 - D_\alpha \Omega_4 - \frac{1}{2} (\gamma^{mn})^\beta_{\ \alpha} D_\beta  \Omega_{5 mn },
\end{equation}
and performing a computation similar to \ref{check_Lambda_trace}, one has
\begin{equation}
     \bar{G}_\alpha + \delta_\Lambda G_\alpha = \frac{1}{2} D_\alpha \Lambda + \frac{1}{4} \gamma^m_{\xi \alpha} \big(  W^{1 \xi } A^2_{m} + \partial_m \Omega^\xi_{2} \big) = 0.
\end{equation}
This is the equation \ref{vanishingofG}. So the vertex operator \ref{answer} has been fixed to the gauge \ref{gauge_fixing_procedure1},\ref{gauge_fixing_procedure2},\ref{gauge_fixing_procedure2}, where it can be written as
\begin{align}
    V^{(12)}_{m^2 = 2} = : \partial \theta^\beta \lambda^\alpha \big( \gamma^{mnp}_{\alpha \beta} B_{mnp} \big): &+ : \Pi^m \lambda^\alpha H_{m \alpha} : + : d_\beta \lambda^\alpha C^{\beta}_{\ \alpha} : + : \frac{1}{2}  N^{mn} \lambda^\alpha F_{ m n \alpha } : , \label{final_answer}
\end{align}
with 
\begin{align}
    B_{mnp} &= \frac{1}{36} W^{1}\gamma_{mnp}W^2 \label{campo_Bfixed} \\ &- \frac{1}{36} ik^{1}_{[m} ik^2_{n} W^{1} \gamma_{p]} W^2 + \frac{1}{72} \partial^r\big(  F^1_{r[m}F^2_{np]} + F^2_{r[m}F^1_{np]}  \big)   \nonumber  \\ 
     H_{m \alpha} &= \frac{3}{7} (\gamma^{st})_{\alpha}^{\beta}D_\beta B_{mst}, \label{campo_Hfixed} \\
    C^{\alpha}_{\beta} &= \frac{1}{4} (\gamma^{ mnpq})^{\alpha}_{\ \beta} \partial_{m}B_{npq},    \label{campo_Cfixed} \\
   F_{mn \alpha} &= \frac{1}{16} \big( 6 \mathcal{H}_{mn \alpha} -  (\gamma_{p [m})_{\alpha}^{\ \beta} \mathcal{H}_{n] p \beta} \big), \ \ \    \mathcal{H}_{mn\alpha} \equiv  \partial_{[m} H_{n] \alpha}\label{campo_Ffixed}
\end{align}
and therefore gives a SYM realization of the massive spin-two multiplet of mass $(mass)^2 = 2$.

\section*{Acknowledgements}
I would like to thank Nathan Berkovits 
for helping me on crucial stages of this work, and Carlos Mafra and Luis Alberto Ypanaqu\'e for reading and commenting on the manuscript. I would also like to acknowledge FAPESP grant 2022/00869-6 for financial support.


\begin{appendices}
\section{ Conventions and Gamma Matrix formulas}

The gamma matrices satisfy
\begin{equation}
    \left(\gamma^m\right)^{\alpha \sigma} \gamma_{\sigma \beta}^n+\left(\gamma^n\right)^{\alpha \sigma} \gamma_{\sigma \beta}^m=2 \delta^{m n} \delta_\beta^\alpha,
\end{equation}
and the antisymmetrization is represented by square brackets, 
\begin{equation}
    \gamma^{m_1 ... m_k} \equiv \frac{1}{k!} \gamma^{[m_1} ... \gamma^{m_k]}  \equiv \frac{1}{k!} \big( \gamma^{m_1} ... \gamma^{m_k} + \text{all antisymmetric permutations} \big). 
\end{equation}
There are the following important identities, 
\begin{align}
\gamma_{\alpha(\beta}^{m} \gamma_{\gamma \delta)}^{m} & =0 \label{gammaid1}\\
\gamma_{\alpha[\beta}^{m n p} \gamma_{\gamma \delta]}^{m n p} & =0 \label{gammaid2}\\
\gamma_{m n p}^{\alpha \beta} \gamma_{\gamma \delta}^{m n p} & =48\left(\delta_{\gamma}^{\alpha} \delta_{\delta}^{\beta}-\delta_{\gamma}^{\beta} \delta_{\delta}^{\alpha}\right) \label{gammaid3}\\
\gamma_{\alpha \beta}^{m n p} \gamma_{\gamma \delta}^{m n p} & =12\left(\gamma_{\alpha \delta}^{m} \gamma_{\beta \gamma}^{m}-\gamma_{\alpha \gamma}^{m} \gamma_{\beta \delta}^{m}\right) \label{gammaid4}\\
\gamma_{\alpha \beta}^{m} \gamma_{\delta \sigma}^{m} & =-\frac{1}{2} \gamma_{\alpha \delta}^{m} \gamma_{\beta \sigma}^{m}-\frac{1}{24} \gamma_{\alpha \delta}^{m n p} \gamma_{\beta \sigma}^{m n p}, \label{gammaid5}\\
\gamma_{\alpha \beta}^{m n p} \gamma_{\delta \sigma}^{m n p} & =-12 \gamma_{\alpha \beta}^{m} \gamma_{\delta \sigma}^{m}-24 \gamma_{\alpha \delta}^{m} \gamma_{\beta \sigma}^{m}, \label{gammaid7}\\
\left(\gamma^{m n}\right)_{\alpha}{ }^{\delta}\left(\gamma_{m n}\right)_{\beta}{ }^{\sigma} & =-8 \delta_{\alpha}^{\sigma} \delta_{\beta}^{\delta}-2 \delta_{\alpha}^{\delta} \delta_{\beta}^{\sigma}+4 \gamma_{\alpha \beta}^{m} \gamma_{m}^{\delta \sigma}, \label{gammaid8}
 \end{align}
\begin{align}
    \gamma^{m_1 ... m_k} &= \gamma^{m_1} \gamma^{m_2 ... m_k} - \frac{1}{(k-2)!} \delta^{m_1 [ m_2} \gamma^{m_3 ... m_k]},\ \  k = 2,...,5;  \label{formid0} \\
    \gamma^m \gamma^{n_1 \ldots n_k} \gamma_m &= (-1)^k(10-2 k) \gamma^{n_1 \ldots n_k},  k = 2,...,5; \label{formid1} \end{align} \begin{align}
    \gamma^{s t} \gamma_{m n p q r} \gamma^{s t} &= 10 \gamma_{m n p q r}, \\
    \gamma^{s t u} \gamma_{m n p q r} \gamma^{s t u } &= 0, \\
    \gamma^{s t u v} \gamma_{m n p q r} \gamma^{s t u v} &= 240 \gamma_{m n p q r},\\
    \gamma^{s t} \gamma_{m n p q } \gamma^{s t} &= 6 \gamma_{m n p q }, \\
    \gamma^{s t u} \gamma_{m n p q } \gamma^{s t u } &= 48 \gamma_{m n p q}, \\
    \gamma^{s t u v} \gamma_{m n p q } \gamma^{s t u v} &= 48 \gamma_{m n p q}, \\
    \gamma^{s t} \gamma_{m n p  } \gamma^{s t} &= -6 \gamma_{m n p} .
\end{align}
The bispinors Fierz decompositions are
\begin{equation}
\chi_\alpha \psi_\beta=\frac{\gamma^m_{\alpha \beta}}{16} \chi \gamma_m \psi+\frac{\gamma^{m n p}_{\alpha \beta}}{3 ! 16} \chi \gamma_{m n p} \psi+\frac{ \gamma^{m n p q r}_{\alpha \beta}}{5 ! 16\cdot 2}\chi \gamma_{m n p q r} \psi, \label{fierzbb}
\end{equation}
\begin{equation}
\chi_\alpha \psi^\beta=\frac{\delta_\alpha^\beta}{16} \chi \psi-\frac{\left(\gamma_{m n}\right)_\alpha^\beta}{2 ! 16}\chi \gamma^{m n} \psi+\frac{\left(\gamma_{m n p q}\right)_\alpha^{\ \beta}}{4 ! 16}\chi \gamma^{m n p q} \psi . \label{fierzbc}
\end{equation}
Trace relations are given by
\begin{equation}
 \operatorname{Tr}\left(\gamma^{m_{1} \ldots m_{k}} \gamma_{n_{1} \ldots n_{k}}\right)=  (-1)^{\frac{k \cdot (k-1)}{2}} 16 \cdot k ! \delta_{n_{1} \ldots n_{k}}^{m_{1} \ldots m_{k}} + \delta^{k 5} 16 \epsilon^{m_{k} \ldots m_{k}}{}_{n_{1} \ldots n_{5}} , \ \ k = 1,..,5.
\end{equation} 

\end{appendices}


\begin{thebibliography}{00}
\bibitem{Berkovits:2000fe}
N.~Berkovits,
JHEP \textbf{04} (2000), 018
doi:10.1088/1126-6708/2000/04/018
[arXiv:hep-th/0001035 [hep-th]].
\bibitem{Berkovits:2002qx}
N. Berkovits and O. Chandia,
JHEP \textbf{08} (2002), 040
doi:10.1088/1126-6708/2002/08/040
[arXiv:hep-th/0204121 [hep-th]].
\bibitem{Berkovits:2000yr}
N.~Berkovits and O.~Chandia,
Nucl. Phys. B \textbf{596} (2001), 185-196
doi:10.1016/S0550-3213(00)00697-0
[arXiv:hep-th/0009168 [hep-th]].
\bibitem{Berkovits:2000nn}
N.~Berkovits,
JHEP \textbf{09} (2000), 046
doi:10.1088/1126-6708/2000/09/046
[arXiv:hep-th/0006003 [hep-th]].


\bibitem{Berkovits:2005ng}
N.~Berkovits and C.~R.~Mafra,
Phys. Rev. Lett. \textbf{96} (2006), 011602
doi:10.1103/PhysRevLett.96.011602
[arXiv:hep-th/0509234 [hep-th]].

\bibitem{Chakrabarti:2017vld}
S.~Chakrabarti, S.~P.~Kashyap and M.~Verma,
JHEP \textbf{01} (2018), 019
doi:10.1007/JHEP01(2018)019
[arXiv:1706.01196 [hep-th]].

\bibitem{Chakrabarti:2018mqd}
S.~Chakrabarti, S.~P.~Kashyap and M.~Verma,
JHEP \textbf{10} (2018), 147
doi:10.1007/JHEP10(2018)147
[arXiv:1802.04486 [hep-th]].
\bibitem{Friedan:1985ge}
D.~Friedan, E.~J.~Martinec and S.~H.~Shenker,
Nucl. Phys. B \textbf{271} (1986), 93-165
doi:10.1016/0550-3213(86)90356-1

\bibitem{1986CMaPh.106..183H} Harnad, J. \& Shnider, S.\ 1986, Communications in Mathematical Physics, 106, 183. doi:10.1007/BF01454971


\bibitem{Kashyap:2023cdi}
S.~P.~Kashyap, C.~R.~Mafra, M.~Verma and L.~A.~Ypanaqu\'e,
[arXiv:2311.12100 [hep-th]].

\bibitem{DiFrancesco:639405}
Di Francesco, Philippe and Mathieu, Pierre and Sénéchal, David, Springer (1997), 10.1007/978-1-4612-2256-9 
\end{thebibliography}






\end{document}